\documentstyle [12pt] {article}

 \newcommand{\cs}[3]{{{#3} \brace {#1 #2}}}

 \newcommand{\h}[1]{\mathop{\lambda}\limits_{#1}\ \!\!\!}

 \newcommand{\m}{\mu}
 \newcommand{\n}{\nu}

\begin{document}

\begin{center}
\bf {TEMPERATURE FLUCTUATION AND AN EXPECTED LIMIT OF HUBBLE
PARAMETER IN THE SELF-CONSISTENT MODEL }
\end{center}

\begin{center}
{A.B.Morcos}
\end{center}

\begin{center}
Department  of Astronomy, National Research Institute \\Of
Astronomy and Geophysics, Helwan,Cairo,Egypt.\\
{\it e-mail: morcos@frcu.eun.eg}
\end{center}

\begin{abstract}
    The temperature gradient of microwave background radiation (CMBR) is
    calculated in the Self Consistent Model. An expected values
    for Hubble parameter have been presented in two different cases.
    In the first case the temperature is treated  as a function of time
    only ,while in the other one the temperature  depends on relaxation of
    isotropy condition in the self-consistent model
    and the assumption that the universe  expands adiabatically. The
    COBE's or WMAP's fluctuations in temperature of CMBR may be used to
    predict a value for Hubble parameter.

\end{abstract}

\section{Introduction}
         ~~~~~~The cosmic micro wave background radiation (CMBR)
     temperature is one of the important parameters of any cosmological model.
     The three characteristics of this radiation are its
     spectrum, spatial anisotropy and polarization. The COBE
     Far-Infrared Absolute Spectrophotometer (FIRAS) has
     determined the black body temperature of (CMBR) to be
     $2.728\pm 0.004 ~K~$(Keating et al. (1998)), and the COBE
     Differential Microwave Radiometer (DMR) experiment has
     detected spatial anisotropy of the (CMBR) on $~10^{o}~$scales
     of $~\frac{\triangle T}{T} \simeq~1.1\times~10^{-5}~$K. Ground
     and balloon-based experiments have detected anisotropy at
     smaller scales (cf. Silk, and White(1995)). Many authors
     referred these anisotropies to simple linear or nonlinear
     processing in the primordial fluctuations (cf. Hu et al. (1997),
     Challinor and Lasenby(1999), Melek(2002)).\\
      $~~~~$   Recently Wilkinson Microwave Anisotropy Probe (WMAP) satellite,
     which is designed for precision measurement of the CMBR
     anisotropy on the angular scales ranging from the full sky down to
     several arc minutes. This ongoing mission has already
     provided a sharp record of the conditions in the universe
     from the epoch of last scattering to the present. WMAP
     results despite the absence of a direct dark energy interaction
     with our baryonic world (Rebort Caldwell and Michael
     Doran(2003)). Whenever Joshue et al.(2003) show that the
     Planck CMBR mission can be significant. In general the
     observational limit of the temperature
     fluctuations$~~\frac{{\Delta}T}{T}~~$becomes lower and lower
     and it reached almost $~10^{-6}~$ (Keating et al. (1998)).\\
     $~ ~~~~$In the present work ,the COBR or WMAP  results for the temperature
     gradient is used to expect a value for the Hubble constant using the
        self-consistent model. In the next section a brief review
        of the self-consistent model will be given. In
     Section 3 time-temperature relation in the
       model is calculated. In  Section 4
        the theoretical technique for calculating the gradient of
        any scalar cosmic field is described. In Section 5
        a lower limit for Hubble parameter is calculated. In
         Section 6 discussion and concluding remarks are given.

\section{The Self-Consistent Model (SCM)}
     ~~~~~Wanas(1989), has constructed  the (SCM), a cosmological model in
     the frame work of Generalized Field Theory (GFT)(Mikhail and Wanas (1977)).
     This theory is constructed in a
     4-dimensional Absolute Parallelism (Ap)-Geometry. In (1986) Wanas suggested a set of
     conditions to be satisfied by any geometric structure in
     (AP-Geometry) to be suitable for cosmological applications.
     This set of conditions, if satisfied, would guarantee that a geometric structure
     would represent, a homogeneous, isotropic, electrically neutral
     and non-empty universe. Wanas (1989) has used one of the AP-structures,
     constructed by Robertson(1932), satisfying the conditions
     mentioned above, to construct (SCM).
        The geometric structure used in that model is given in
     the spherical polar coordinates by,
     $$ \h{i}^{\mu} ={\left(\matrix{
  \sqrt{-1} & 0 & 0 & 0 \cr
  0 & \frac{L^+ sin\theta cos\phi}{4 R} & \frac{(L^-cos\theta
cos\phi-4K^{\frac{1}{2}}r sin\phi)}{4 r R} & \frac{-(L^-
sin\phi+4k^{\frac{1}{2}}r cos\theta cos\phi)}{4rR  sin\theta}\cr
0 & \frac{L^+ sin\theta sin\phi}{4 R}  & \frac{(L^- cos\theta
sin\phi-4K^{\frac{1}{2}}r cos\phi)}{4 r R} & \frac{(L^-
cos\phi-4k^{\frac{1}{2}}r cos\theta sin\phi)}{4rR sin\theta}\cr  0
& \frac{L^+ cos\theta}{4 R} & \frac{-L^- sin\theta}{4 r R} &
\frac{K^{\frac{1}{2}}}{4 R} }\right)}~~.\eqno{(1)}$$ where $L^\pm
= 4 \pm k r^{2},$ k is the curvature of the space and $R(t) $ is
an unknown function of (t) only.
 It is to be considered that the Riemannian space, associated with(1), is given by\\
  $$dS^{2}=\hat{g}_{\m\n}~dx^{\mu}~dx^{\nu}.\eqno{(2)}$$
  with the metric tensor given by,
  $$\hat{g}_{\m\n}=\sum_i  e_{i}\h{i}_\mu~\h{i}_{\nu},$$
  $$\hat{g}^{\m\n}{\stackrel{def.}{=}}\sum_i e_{i}\h{i}^\mu~\h{i}^{\nu},\eqno{(3)}$$
  where $e_{i}(=1,-1,-1,-1)$ is Levi-Civita's indicator.
  Wanas(1989) got the following set of the differential equations:
  $$\frac{\dot{R^{2}}}{R^{2}}+\frac{4k}{R^{2}}=0,\eqno{(4)}$$

$$\frac{2\ddot{R}}{R}+\frac{\dot{R^{2}}}{R^{2}}+\frac{4k}{R^{2}}=0,\eqno{(5)}$$
where the dots represents differentiation with respect to time
(t). Integration of (4), gives immediately
$$R=\tilde{R}\pm~2(-k)^{\frac{1}{2}}~t,\eqno{(6)}$$ where
$\tilde{R}$ is a constant of integration, giving the value of the
scale factor  at t=0. If k takes the value zero , the SCM model
will be a static empty one,and when k=+1 it will give an imaginary
scale factor. So we must take k=-1 for non-static, non-empty
model,and the solution (6) will take the form,
$$R=\tilde{R}+2t,\eqno(7)$$
with k = -1. It is worth of mention that the SCM is a cosmological
model fixing the the curvature constant to be -1. It is also
satisfying the weak and strong energy conditions, and it is free
of particle horizon( for more details see Wanas(1989), (2003)).
The model is consistent with recent Supernovae observations  Riess
et al. (2004). The negative curvature, uniquely fixed by the
model, is among recent discussed reasons of the WMAP low
multi-pole
 anomaly (cf. Gurzadyan et al. (2003)). So, this model deserves
 further examination.

\section{The Time-Temperature Relation in SCM}

    ~~~~~In what follows we are going to find the relation between
    time and temperature in SCM, we are going to assume that in
     the early stages of the universe, the radiations behaves as if it is
     coming from a black body with temperature T given by the will known
    relation(cf. Narlikar (1983))
   $$B_0~^0~ = a~ T^4,\eqno(8)$$
    where $B_\m~^{\n}$ is the phenomenological energy-momentum
    tensor, and $~a~$is the radiation constant. But the energy
   momentum tensor in the SCM is a geometric one, say $S_\m~^{\n}$,
that has the non-vanishing values:
$$S_{0}~^{0}={\frac{9k}{R^{2}}}~,~S_{1}~^{1}=S_{2}~^{2}=S_{3}~^{3}=\frac{3k}{R^{
2}}.\eqno(9)$$We can assume that the geometric energy momentum
tensor is related to the phenomenological one via the relation,
$$S_\m~^{\n}={\mathcal{H}} {B_\m~^{\n}},\eqno(10)$$
where  $\mathcal{H}~$ is a conversion constant equal to
$\frac{8~\Pi~G}{c^{2}}$, G is the gravitational constant and c is
the speed of light. If we use (7), (8), and (9), we get
$$ T=(\frac{9}{4a\mathcal{H}})^{1/4}~(\frac{2}{\tilde{R}+2t})^{1/2}.\eqno(11)$$
But it is well known that the relation between temperature and
time depends on the type of particles filling the model and the
kind of interaction between them at a certain temperature range.
Thus it is more convenient to rewrite the relation(11) in the
form,
$$ T=(\frac{9}{a{\mathcal{H}}~\gamma~(\tilde{R}+2t)^{2}})^{1/4}~,\eqno(12)$$
where $~\gamma~$ is a parameter depending on types of particles
and their interactions. The relation (12) may be used to determine
the parameter$~\tilde{R}~$ , if all other constants are known.
$~~~~$If it is assumed that at time t=0 the temperature of the
 universe is $~10^{12}~^{o}K$, as it is usually used in the thermal
 literature (cf. Narlikar(1983)), the value of the parameter$~\tilde{R}~$ is
 obtained from the relation (12) to be $~3.7\times10^{-4}sec~$. Relation (12) then
 takes the form $$T=(\frac{9}{a{\mathcal{H}}~\gamma~(3.7\times10^{-4}+2t)^{2}})^{1/4}~,\eqno(13)$$
where  $~\gamma = 1.45~$ (cf. Narlikar (1983)).
\section{The Gradient of any Cosmic Scalar Field}

          ~~~~~ Melek(1992) generalized a procedure, used in
           meteorology,
      in studying the temperature gradient in the Earth's
      atmosphere, to study the matter density and temperature gradients in the
       universe. For any cosmic measurable scalar field S which
      can be related to the energy momentum tensor $
      B_{\mu\nu}$, he defined the function $F_{g}$, in a curved
      space-time with metric $g_{\mu\nu}$, as :

      $$F_{g}{\stackrel{def.}{=}}\frac{dG}{d\tau},\eqno(14)$$
      $$where~~~~G{\stackrel{def.}{=}}(g^{\mu\nu}S_{\mu} S_{\nu})^{1/2},\eqno(15) $$
      $$and ~~~~S_{\mu} =\frac{\partial S}{\partial
      x^{\mu}},\eqno(16)$$
      where $S_{\mu}$~~is a time-like covariant vector, $\tau$~~is the
      cosmic time and $~\mu~=0,1,2,3$.
      Melek has shown that the function $F_{g}$ has the form:
      $$F_{g} ={\frac{1}{G}}~g^{\mu\nu}
      ~S_{\mu;\sigma}~S_{\nu}~u^{\sigma},\eqno(17)$$
      where $~S_{\mu;\sigma}~$ is the usual covariant derivative
      with respect to
      $~x^{\sigma}~~$and~\\$u^{\sigma}{\stackrel{def.}{=}}{\frac{dx^{\sigma}}{d\tau}}~$.
     ~The second derivative of the absolute value of the
      gradient of any cosmic scalar field S, with respect to the
      cosmic time ~$\tau$~, is given by:
      $$\frac{d^{2}G}{d\tau^{2}}= \frac{dF_{g}}{d\tau}=
      {\frac{1}{G}}~\{{g^{\mu\nu}u^{\sigma}u^{\alpha}[
      ~S_{\mu;\sigma\alpha}~S_{\nu}+
      S_{\mu;\sigma}S_{\nu;\alpha}]-F^{2}_{g}}\}.\eqno(18)$$
      ~~~~~Melek(1995) applied this procedure to suggest an expression
      for the function ~$F_{g}$~~in a spatially perturbed
      Fredman-Robertson-Walker cosmological model(FRW). He put a lower
      limit on the Hubble parameter. Melek (2000) used the same
      technique for (FRW) to study limits on cosmic time scale
       variations of
      gravitational and cosmological constants. Melek(2002)used the same procedure
      to find the primordial angular gradients in the temperature of the
      microwave background radiation and the density functions in the same
      cosmological model.

    In what follows we are going to use the same technique
      to find the gradient of microwave back ground radiation's temperature
      in SCM . Also, we are going to get a relation between this
      gradient and the value of Hubble parameter.

      \section{ CMBR Temperature Gradient in the SCM and
      Expected Limit of Hubble Parameter}

           The metric of the Riemannian space, associated with the AP-space (1),
            can be  written using, equations (2) and (3), as

       $$dS^{2}= d t^2 - {\frac{16~R^2}{{L^{+}}^2}}[{dr^2 +~r^2~d{\theta}^2+~r^2~sin
           {\theta}~d{\phi}}]
           ,\eqno(19)$$
           where $~L^+ = 4+r^2k$.\\
Now if we follow the coordinate transformation, \\
$$dt=R(t)d\tau\\,\eqno(20)$$
  in the metric (19),then we can write
$$dS^{2}= R^2(t)\{{d
{\tau}^2} - {\frac{16}{{L^{+}}^2}}[{dr^2 +
           ~r^2~d{\theta}^2+
           ~r^2~sin
           {\theta}~d{\phi}}]\}
           ,\eqno(21)$$
where R(t) is the scale factor and   $\tau$ is the cosmic time. If
we assume that the microwave background radiation temperature
(T(t)), is our scalar field and this field varies with time only,
then following (16), we can write
$$ T_\mu{\stackrel{def.}{=}}\frac{d~T}{d~x^\mu},\eqno(22)$$

$$T_0=\frac{d~T}{d~t}~\frac{d~t}
{d~\tau}\\=R~\stackrel{.}{T},\eqno(23)$$ where
$\stackrel{.}{T}=\frac{d~T}{d~\tau}$~. Then,
$$T_{\mu~;~\nu}=T_{\mu~,~\nu}-\cs{\mu}{\nu}{\rho}~T_{\rho}.\eqno(24)$$

Now by taking into consideration that the temperature is a
function of time only and using (24),we can write
$$T_{0~;~0}= R(R~\stackrel{..}{T}-\stackrel{.}{R}~\stackrel{.}{T}).\eqno (25)$$
 Using equations (14),(15),(16),(17),(18)and (25),taking into consideration
 that the CMBR is independent of the radial coordinate at any fixed cosmic time
 and the motion in the Universe is only due to its expansion, then we get after some
 straight forward calculations ,
 $$F=~(\stackrel{..}{T}-\frac{\stackrel{.}{R}}{R}~\stackrel{.}{T}).\eqno (26)$$
  Since the SCM has been assumed to be homogenous and isotropic then
 $~F=0 ~$ i.e,
$$\stackrel{..}{T}-\frac{\stackrel{.}{R}}{R}~\stackrel{.}{T}~=~0.\eqno (27)$$
Equation (27) leads directly to the following result
$$\frac{\stackrel{..}{T}}{\stackrel{.}{T}}=\frac{\stackrel{.}{R}}{R}~.\eqno (28)$$
Noting that$~\frac{\stackrel{.}{R}}{R}= H~$, as usually done, then
we get
$$\frac{\stackrel{..}{T}}{\stackrel{.}{T}}=H~.\eqno (29)$$

  It is clear from the last equation that all the quantities on
  its left hand side are unmeasurable quantities till now, so if
  these quantities are measured by COBE, WMAP or any other satellite, the
  Hubble parameter is determined completely and at that moment can
  be fixed.

  As it is mentioned above, the most recently  detected value of
  the anisotropy in the temperature of the CMBR is determined by COBE
  and WMAP for each $~10 ^{o}~$.
   This means that it is more suitable to relax the
  condition of isotropy  in the cosmological model used. To
  satisfy this aim we are going to use the spatially perturbed form of
  the metric of the SCM in the spherical polar coordinates. The
  metric (21) will take the form:

    $$dS^{2}= R^2(t)\{{d
    {\tau}^2} - {\frac{16(1+h_{1})}{{L^{+}}^2}}{dr^2 }-
    h_{2}~r~dr~d\Omega~-~r^{2}~(1+h_{3})~d\Omega^{2}\},\eqno (30)$$
    where $~\Omega~$ is ~the solid ~angle ~defined ~in terms of
    ~$~\theta~$ ~and~ $~\phi~$ as~ $
    d\Omega^{2}~=~d\theta^{2}~+~\sin^{2}\theta~d\phi^{2}$
    and $h_{1},h_{2}~and~ h_{3}~$ are small spatial
    perturbations. If we use the metric (30) taking into our
    consideration that the homogeneity is valid (i.e $
    \frac{\partial~T}{\partial~r}~=~0~, and~
    \frac{\partial^{2}~T}{\partial~t~\partial~r}~=~0$ ), and assuming
    that the expansion is the only motion in the universe,then this expansion
    affects the temperature. If we assume now that the temperature of the CMBR is a
    function of the cosmic time and direction i.e $T( t, \Omega)$ and one
    follows the same procedure as before, then equation (17) takes the
    form
    $$F= ({\frac{1}{G~R(t)}})(\stackrel{.}{T})(\stackrel{..}{T}-
    \frac{\stackrel{.}{R}}{R}~\stackrel{.}{T})
    -(\frac{1-h_{2}}{r^{2}})({T'})(\frac{\partial{T'}}{\partial~t}
    -\frac{\stackrel{.}{R}}{R}~{T'}),\eqno (31)$$
      where$~~{T'}~{\stackrel{def.}{=}}~ \frac{\partial~T}{\partial~\Omega}$.
      If we write now the metric of the SCM , which is homogenous and
      isotropic,in the form
          $$dS^{2}= R^2(t)\{{d
          {\tau}^2} - {\frac{16}{{L^{+}}^2}}{dr^2 }
          -~r^{2}~d\Omega^{2}\},\eqno (32)$$
     then, after some straight forward calculations, we can find the temporal
     variation of the magnitude of the gradient of T as
      $$F_{SCM}= ({\frac{1}{G~R(t)}})(\stackrel{.}{T}){(\stackrel{..}{T}-
      \frac{\stackrel{.}{R}}{R}~\stackrel{.}{T})}
      -({\frac{1}{r^{2}}})({T'})({\frac{\partial{T'}}{\partial~t}
      -\frac{\stackrel{.}{R}}{R}~{T'})}.\eqno (33) $$
      This gradient will be zero if the model is homogeneous and
      isotropic. From
      equations (31) and (33), assuming that the universe expands
      adiabatically, we get
     $$({\frac{\partial{T'}}{\partial~t}
     -\frac{\stackrel{.}{R}}{R}~{T'})}~=~0.\eqno (34) $$
     Since $~H~= \frac{\stackrel{.}{R}}{R}~$, then the Hubble
     parameter can be written as
     $$~ H~=~(\frac{\partial{T'}}{\partial~t})~/~{{T'}}.\eqno (35)$$
 Since COBE, WMAP and other space and ground based measurements have
 detected  and confirmed anisotropy in the temperature of the CMBR,
 this means that the right hand side quantities of the
 equation (35), can be measured easily fixing the value
 of the Hubble parameter.
    \section{Discussion and Concluding Remarks}
Using theAP-structure (1), Wanas(1989) has got a unique pure
geometric world model . This model is non-empty and has no
particle horizons. This model fixes a value for $k(=-1)$ i.e. it
has no flatness problem, and as it is clear from equation (7), it
has no singularity at t=0.
 A further advantage of using pure geometric theories is that
 one did not need to impose any condition from outside the geometry
 used (e.g. equation of state) in order to solve the field equations
 (Wanas (1986) , (1989)).\\

      In the present work the generalized procedure for studying gradients,
      which has been used by Melek (1992),
 is used to find the temperature gradient in the SCM. It is
 shown that when it is
assumed that the CMBR temperature is a function of time only, the
Hubble parameter (H) is given by (22). But all the quantities on
the right hand side are nonmeasurable, so this relation can not
determine the numerical value of H except for a satellite or a
ground based observation arises the gradient of temperature and
its
rate of change with respect to time. \\

When the isotropy condition in the self-consistent model is
relaxed and the universe is assumed to expand adiabatically, the
Hubble parameter is given by the relation (35). The quantity in
the denominator of the right hand side of (35) may be determined
by COBE or WMAP while the quantity in the numerator can not be
determined at time being. It can be calculated
after the accumulation of further data, and then the Hubble parameter can be determined.\\

It is clear also from the relation (34) that the value of the
Hubble parameter decreases as the temperature gradient decreases.
This result is in agreement with Bellini(2001) results.\\

It is worth of mention that the gradient relation my give the same
form for many of cosmological models but each result depends
essentially on the value of the scale factor fixed by the model
under consideration, i.e this procedure is model dependent.
 \section{Acknowledgement}
 The author would like to express his deep thankfulness to
 Professor M.I.Wanas for his useful discussions.Also he is
 indebted to the late Dr. M.Melek for his previous guiding
 points. Part of this work has a been accomplished during the author's
 visit to the High Energy Section of the Abdos Sallam ICTP. So, the author
 would like to thank Professor S. Randjbar-Daemi
 the Head of High Energy Section, ICTP, Italy, for inviting him to
 visit and use the facilities of the ICTP, during the period from 25
  July to 3 August 2004.
      \section{References}

 Bellini, M. (2001) GRG, {\bf 33}, 2081.\\
Challinor, A. and Lasenby, A.(1999)Ap. J., {\bf 512}, 1.\\
Gurzadyan V.G. et al. (2003), Astro-ph/0312305.\\
Hu,W., Sugiyama, N. and Silk, J.(1997) Nature, 386.\\
Joshue , A. F., Dragan H., Eric, V. L. , Michael,S. T.(2003)

astro-ph/0208100v2.\\
Keating, B., Timbie, P., Polnarev, A. and  Steinberger, J.(1998)

 Ap. J., {\bf 495}, 580.\\
Melek, M.(1992) ICTP print no. IC/92/95.\\
Melek, M.(1995) Astrophys. Space. Sci., {\bf 228}, 327. \\
Melek, M.(2000) Astrophys. Space. Sci., {\bf 272}, 417. \\
Melek, M.(2002) Astrophys. Space. Sci., {\bf 281}, 743. \\
Mikhail, F.I. and Wanas, M.I. (1977) Proc. Roy. Soc. Lond.
A,356,471.\\
Narlikar (1983), Introduction to Cosmology, Jones and Bartlett

Publishers, Inc.\\
Rebort, R. C. and Doran, M. (2003), astro-ph/0305334v1.\\
Robertson, H. P. (1932) Ann. Math. Princeton(2),{\bf 33},496.\\
Riess, A. G. , et al. (2004)Ap. J., {\bf 607}, 665.\\
 Wanas, M. I. (1986) Astrophys. Space. Sci., {\bf 127}, 21. \\
 Wanas, M. I.  (1989) Astro. Space Sci., {\bf 154}, 165.\\
 Wanas, M. I. (2003) Chaos, Solitons and Fractals, {\bf 16}, 621.\\
\end{document}